\magnification 1200

\font\Bbb=msbm10
\def\BBB#1{\hbox{\Bbb#1}}

\font\Frak=eufm10
\def\frak#1{{\hbox{\Frak#1}}}

\font\small=cmr8

\def\B{{\bf B}}
\def\d{\partial}
\def\bv{{\bf v}}
\def\w{{\bf w}}
\def\div{{\rm div}}
\def\curl{{\rm curl}}
\def\ad{{\rm ad}}
\def\Id{{\rm Id}}

\def\R{\BBB R}

\def\SV{{\rm SVect}}
\def\SVD{\SV(D)}
\def\SD{{\rm SDiff}}
\def\SDD{\SD(D)}
\def\K{\Omega^1 / d \Omega^0}
\def\KD{\Omega^1(D) / d \Omega^0(D)}
\def\g{{\frak g}}
\def\tv{{\widetilde \bv}}
\def\tB{{\widetilde \B}}
\def\utB{{\widetilde B}}
\def\utv{{\widetilde v}}

\def\Alfven{{Alfv{\c{e}}n }}

\centerline{\bf Magnetic hydrodynamics with asymmetric stress tensor.}

\

\centerline {\bf Yuly Billig}
\centerline{School of Mathematics and Statistics, Carleton University,}
\centerline{1125 Colonel By Drive, Ottawa, K1S 5B6, Canada}
\centerline{e-mail: billig@math.carleton.ca} 

\

\

{\narrower\noindent\small
% {\bf Abstract}:
In this paper we study equations of magnetic hydrodynamics with a
stress tensor. We interpret this system as the generalized Euler
equation associated with an abelian extension of the Lie algebra
of vector fields with a non-trivial 2-cocycle. We use the Lie algebra
approach to prove the energy conservation law and the conservation 
of cross-helicity.
\par}

\

\

{\bf 1. Introduction.}

Magnetic hydrodynamics (MHD) describes evolution of a fluid or plasma,
carrying a magnetic field. This theory is used to model the processes
in the Solar corona [17], as well as to design tokamaks [10]. There are numerous
books treating various aspects of the subject, see e.g., [8], [15].

The MHD equations are derived from the Euler equation of motion of an 
incompressible fluid and the Maxwell's electrodynamics equations, and 
describe evolution of a fluid with the velocity vector field $\bv$ 
and the magnetic field $\B$:
$$\left\{
\matrix{
{\d \bv \over \d t} = - (\bv \cdot \nabla) \bv + (\B \cdot \nabla) \B 
- \nabla p, \hfill \cr
{\d\B \over \d t} = - \left\{ \bv, \B \right\}, \hfill \cr
\div (\bv) = 0, \, \, \div (\B) = 0. \hfill \cr
}
\right.
\eqno{(1.1)}
$$

In this paper we study another system of PDEs, where we add extra terms
into the evolution equation for the velocity:
$$ 
{\d \bv \over \d t} = - (\bv \cdot \nabla) \bv + (\B \cdot \nabla) \B 
+ \sum\limits_{i,j} {\d B_i \over \d x_j} \nabla {\d v_j \over \d x_i} 
- \nabla p .
\eqno{(1.2)}
$$
The additional terms here are of the third degree in derivatives and one
can draw certain parallels between (1.2) and the Korteweg - de Vries equation 
$$ u_t = u u_x + u_{xxx} . \eqno{(1.3)}$$

We interpret the extra terms in (1.2) as a contribution of a stress tensor
$$ T_{ki} = \sum\limits_j 
{\d B_i \over \d x_j} {\d v_j \over \d x_k} . \eqno{(1.4)}$$
This stress tensor is not symmetric, which indicates that the particles of 
the fluid should possess electric or magnetic momentum.

 In spite of the additional higher-order terms, the new system of PDEs retains many
features of the original system. In particular, we show that it still admits
the \Alfven wave solutions.

We follow an approach developed by Arnold to give an interpretation of the 
MHD equations with the stress tensor as a generalized Euler equation.
A generalized Euler equation is an equation for the geodesics on a 
(possibly infinite-dimensional) Lie group supplied with a Riemannian structure.
This equation describes the evolution of a tangent vector of the geodesic in the 
Lie algebra of the Lie group (see Section 3 for details).

The Lie algebra that corresponds to the MHD equations with the stress tensor is an 
abelian extension $\g(\tau)$ of the Lie algebra of the divergence zero vector fields 
twisted with a non-trivial 2-cocycle $\tau$.
This Lie algebra was studied in the framework of the representation theory
of the toroidal Lie algebras and the cocycle plays a prominent role there.

Infinite-dimensional groups associated with abelian extensions of the Lie
algebra of vector fields are discussed in [7].

The Lie algebra $\g(\tau)$ has nice properties and this translates into
nice properties of the PDEs. In particular we establish the energy conservation law 
and the cross-helicity conservation for MHD with the stress tensor.

Ovsienko and Khesin showed in [16] that the generalized Euler equation for the
Lie algebra of vector fields on a circle yields the non-linear wave equation
$ u_t = u u_x ,$
while incorporation of the Virasoro cocycle into the Lie algebra leads to
the Korteweg - de Vries equation (1.3). In a way, the present paper may be viewed
as a higher-dimensional generalization of [16].

The paper is organized as follows: in Section 2 we discuss the properties of the
system (1.2), derive the expression for the stress tensor, list the conservation laws
and describe the \Alfven wave solutions. In Section 3 we review the generalized
Euler equation for an arbitrary Lie algebra and we apply this method in Section 4 
to an abelian extension of the Lie algebra of the divergence zero vector fields, deriving
(1.2), and establishing the conservation laws in a purely algebraic way.

\

{\bf 2. Magnetic hydrodynamics with a stress tensor and its properties.}

\

Evolution of an incompressible fluid carrying a magnetic filed 
is given by the equations of magnetic hydrodynamics (MHD):
$$\left\{
\matrix{
{\d \bv \over \d t} = - (\bv \cdot \nabla) \bv + (\B \cdot \nabla) \B 
- \nabla p, \hfill\cr
{\d\B \over \d t} = - \left\{ \bv, \B \right\}, \hfill\cr
\div (\bv) = 0, \, \, \div (\B) = 0. \hfill\cr
}
\right.
\eqno{(2.1)}
$$
Here $\B$ is the magnetic field,
$\bv$ is the velocity vector field of the fluid and $p$ (pressure) is 
an auxiliary function which is chosen in such a way that the equation
$\div (\bv) = 0$ is satisfied. The formal dot product $\bv \cdot \nabla$
represents the differential operator
$$\bv \cdot \nabla = \sum\limits_j v_j {\d \over \d x_j} .$$
The Poisson (Lie) bracket 
$\left\{ \bv, \B \right\}$ of two vector fields  
$\bv = \mathop\sum\limits_j v_j(x) {\d \over \d x_j}$ and 
$\B = \mathop\sum\limits_j B_j(x) {\d \over \d x_j}$ is given by
$$\left\{ \bv, \B \right\} = \ad (\bv) \B = 
(\bv \cdot \nabla) \B - (\B \cdot \nabla) \bv =
\sum\limits_j \left( v_j {\d \B \over \d x_j} - B_j {\d \bv \over \d x_j} 
\right) .$$  
In the three-dimensional space, the first equation in (2.1) may be 
also written as
$${\d \bv \over \d t} =  \bv \times \curl (\bv)  + \curl (\B) \times \B 
- \nabla p .$$
For a conducting medium,
the curl of the magnetic field is equal to the electric current (Amp\`ere's law), 
and so the 
expression $\curl (\B) \times \B$ represents the Lorentz force, with which the
magnetic field acts on the current. 

 The configuration space for a flow of an incompressible fluid is the group
of volume-preserving diffeomorphisms.
 In his remarkable paper [2], Arnold interpreted the Euler equation for an
ideal fluid as the geodesic equation on this Lie group.
The geodesic equation describes the evolution of the tangent vector of the 
geodesic curve, and this tangent vector belongs to the Lie algebra, which is 
the Lie algebra $\SV$ of the divergence zero vector fields, in case of the 
group of volume preserving diffeomorphisms.

 Using Arnold's method, Vishik and Dolzhanskii [19]
(see also [12]) showed that the 
MHD equations (2.1) also may be interpreted as a geodesic equation for a certain
infinite-dimensional Lie group. The Lie algebra that is used to write this equation 
is the semidirect product 
$$\SV \oplus \K$$
of the Lie algebra of divergence zero vector fields $\SV$ with its dual space 
-- the factor
$\K$ of the differential 1-forms modulo the exact 1-forms. 
We review this construction in detail in Section 3.

The Lie algebra that is associated with the MHD equations,
$$\g = \SV \oplus \K ,$$
has recently attracted much interest in representation theory. It turns out that
representations of $\g$ may be used for constructing modules for toroidal Lie algebras
(see e.g., [4], [5], [9]). It was also discovered that this Lie algebra has an important
deformation -- the Lie bracket in $\g$ may be twisted with a Virasoro-like 2-cocycle
$\tau$. The twisted Lie algebra $\g(\tau)$ still has nice properties, and its 
representation theory is even better than of $\g$ itself.

 In this paper we study the system of PDEs that comes from the geodesic equation
for $\g(\tau)$. In Section 4 we show that this Lie algebra 
yields the following system of PDEs:
$$\left\{
\matrix{
{\d \bv \over \d t} = - (\bv \cdot \nabla) \bv + (\B \cdot \nabla) \B 
+ \sum\limits_{i,j} {\d B_i \over \d x_j} \nabla {\d v_j \over \d x_i} 
- \nabla p,  \hfill\cr
{\d\B \over \d t} = - \left\{ \bv, \B \right\}, \hfill\cr
\div (\bv) = 0, \, \, \div (\B) = 0. \hfill\cr
}
\right.
\eqno{(2.2)}
$$
Let us discuss the properties of this system of PDEs.

First of all, we note that the new term
$$\mathop\sum\limits_{i,j} {\d B_i \over \d x_j} \nabla {\d v_j \over \d x_i}
= \mathop\sum\limits_{i,j} {\d^2 \over \d x_i \d x_j} \left( B_i \nabla v_j \right)
\eqno{(2.3)}$$
can also be written as 
$$- \mathop\sum\limits_{i,j} {\d v_j \over \d x_i} \nabla {\d B_i \over \d x_j},
\eqno{(2.3^\prime)}$$
since the difference of the two expressions is the gradient of
$\mathop\sum\limits_{i,j} {\d v_j \over \d x_i} {\d B_i \over \d x_j}$  
and may be absorbed into $\nabla p$. 

It is curious to note here that in the one-dimensional case the passage from the Lie algebra
of vector fields to the Virasoro algebra leads to the transition from the 
non-linear wave equation 
$$ u_t = u u_x $$
to the Korteweg-de Vries equation
$$ u_t = u u_x + u_{xxx},$$
as shown by Ovsienko and Khesin in [16].

Just like the dispersion term $u_{xxx}$ in the KdV, the new term 
$\mathop\sum\limits_{i,j} {\d B_i \over \d x_j} \nabla {\d v_j \over \ d x_i}$
that we get in (2.2) has a triple derivative in $x$.

 Next we are going to show that (2.2) describes magnetic hydrodynamics with
a {\it stress tensor}. Indeed, for a stress tensor $T_{ki}$, the equations 
on the velocity field in (2.1) will become (see e.g. Section 1.7 in [18]) 
$${\d v_k \over \d t} = - (\bv \cdot \nabla) v_k + (\B \cdot \nabla) B_k 
+\sum\limits_i {\d T_{ki} \over \d x_i} - {\d p \over \d x_k}  . \eqno{(2.4)}$$

{\bf Proposition 1.} The system (2.2) describes magnetic hydrodynamics with a
stress tensor. The stress tensor $T_{ki}$ may be written as 
$$ T_{ki} = \sum\limits_j 
{\d B_i \over \d x_j} {\d v_j \over \d x_k} \eqno{(2.5)}$$
or as 
$$ T^\prime_{ki} = - \sum\limits_j 
{\d v_i \over \d x_j} {\d B_j \over \d x_k} \eqno{(2.5^\prime)}$$
or as a linear combination $\alpha T_{ki} + \beta T^\prime_{ki}$ with
$\alpha + \beta = 1$.

\noindent
{\sl Proof.} We will prove the statement of the Proposition for tensor $T_{ki}$ given
by (2.5). We write the contribution in (2.4) from the stress tensor (2.5):
$$ \sum\limits_i {\d T_{ki} \over \d x_i} = 
\sum\limits_{i,j} {\d^2 B_i \over \d x_i \d x_j} {\d v_j \over \d x_k}
+ \sum\limits_{i,j} {\d B_i \over \d x_j} {\d^2 v_j \over \d x_i \d x_k} .$$
Since $\div (\B) = 0$, the first term in the right hand side vanishes, and
we get precisely the first equation from (2.2).  

The stress tensor (2.5${}^\prime$) will yield the additional term written 
in the form (2.3${}^\prime$). The proof in this case is completely analogous.

\

 We point out that the stress tensors we obtain here are not symmetric:
$T_{ki} \neq T_{ik}$. Asymmetric stress tensors occur when the particles
of the fluid are polar, i.e., possess electric or magnetic momentum 
(see Chapter 8 in [18]). The derivation of the stress tensor from the first
principles is rather delicate (see e.g., Chapters 7 in [18]),
and I am unable to describe precisely the physical situations when
the stress tensors (2.5) or (2.5${}^\prime$) would occur.

 Introduction of the term (2.3) which has the 3rd order in derivatives, 
into the equations
will clearly change the behaviour of the solutions in a substantial way.
I view of that, it is quite surprising that the conservation laws of
magnetic hydrodynamics still hold for the system (2.2).

Our next goal is to study the conservation laws for magnetic hydrodynamics with
the stress tensor (2.2). However before we do that, let us discuss the class of 
solutions that we consider here.

 We require that the functions $v_i (x), B_i(x)$ are defined in a domain
$D \subset \R^n$ and belong to the intersection 
of the Sobolev spaces
$\mathop\cap\limits_{k=1}^\infty H_0^k (D)$. 
We recall that the space $H_0^k (D)$ is the closure in the Sobolev space 
$H^k (D)$ of the functions of class $C^k$ with compact support (see e.g. [1]).
This will ensure that all functions and their partial derivatives of all orders
are square-integrable (belong to $L^2 (D)$) and satisfy the vanishing conditions on the
boundary of $D$ (if $D$ is unbounded this means that at infinity the functions go 
to zero faster than the inverse of any polynomial).     
 Choosing this class of functions will allow us to carry out integration by parts with
the boundary term vanishing.

Alternatively, we may consider periodic boundary conditions, as it is often done in 
turbulence theory.

 There are several conserved quantities for the MHD system -- 
mass, momentum, magnetic helicity, as well as energy
and cross-helicity. It turns out that all of these are also
conserved for the MHD with the stress tensor. 
Since we consider the case of an incompressible fluid, the conservation of mass
(volume) holds trivially. The derivation of the conservation of magnetic 
helicity involves only the evolution equation on magnetic field $\B$, and is exactly
the same for both systems. Let us prove the conservation of momentum for the
new system.

{\bf Proposition 2.} The total momentum
is a conserved quantity for the MHD system with the stress tensor (2.2):
$$ \int_D \bv(x) dV \equiv Const .$$

{\sl Proof.}
With the help of Proposition 1, we can write the first equation using the divergence
operator:
$$ {\d v_k \over \d t} = \sum\limits_i 
{\d \over \d x_i} \left( - v_i v_k + B_i B_k + T_{ki} - p \delta_{ki}
\right) .$$

By the divergence theorem, we get
$$ {\d \over \d t} \int_D v_k (x) dV =
 \oint_{\d D} R_{ki} {\bf e}_i \cdot d {\bf n} ,$$
where $R$ is a 2-tensor
$$R_{ki} = - v_i v_k + B_i B_k + T_{ki} - p \delta_{ki} ,$$
${\bf n}$ denotes the unit outward normal vector, and ${\bf e}_i$'s are the
standard basis vectors.
Since the vector fields we consider vanish on the boundary
of $D$, the last integral is zero.

\

Next we state the corresponding theorem
for the conservation of energy and the cross-helicity conservation:

{\bf Theorem 3.} The system (2.2) of magnetic hydrodynamics with stress tensor (2.5)
in a domain $D \subset \R^n$ with appropriate boundary conditions (see discussion 
above)
has the following two first integrals:
$$\int_D \sum_i v_i(x)^2 + \sum_i B_i(x)^2 dV \equiv Const \; \; \; \;
\hbox{\rm (energy conservation)} \eqno{(2.6)} $$
and
$$\int_D \sum_i v_i(x) B_i(x) dV \equiv Const \; \; \; \;
\hbox{\rm (cross-helicity conservation)}. \eqno{(2.7)} $$

We will give the proof of this theorem in Section 4. These conservation laws
will be derived from the properties of the Lie algebra $\g(\tau)$.

It is interesting to note that unlike the case of the Navier-Stokes equation,
introduction of the stress tensor in (2.2) does not lead to the dissipation
of energy, and the energy conservation law still holds.

 For the topological interpretation of helicity and cross-helicity, see [13],
[14].

In conclusion of this section we are going to show that system (2.2) admits \Alfven
wave solutions.

\Alfven waves solutions are obtained as a perturbation of a steady-state constant 
solution $\bv(x) = 0, \, \B(x) = \B_0$. If we take an expansion $\bv = \tv (x), \, 
\B = \B_0 + \tB(x)$ near this equilibrium state, we will get the following system:
$$\left\{
\matrix{
{\d \tv \over \d t} = - (\tv \cdot \nabla) \tv 
+ (\B_0 \cdot \nabla) \tB + (\tB \cdot \nabla) \tB 
+ \sum\limits_{i,j} {\d \utB_i \over \d x_j} \nabla {\d \utv_j \over \d x_i} 
- \nabla p,  \hfill\cr
{\d\tB \over \d t} = - (\tv \cdot \nabla) \tB 
+ (\B_0 \cdot \nabla) \tv + (\tB \cdot \nabla) \tv, \hfill\cr
\div (\tv) = 0, \, \, \div (\tB) = 0. \hfill\cr
}
\right.
\eqno{(2.8)}
$$
Next we set $\tB = \tv$. In this case the term 
$\sum\limits_{i,j} {\d \utB_i \over \d x_j} \nabla {\d \utv_j \over \d x_i}$
is a gradient of 
$\sum\limits_{i,j} {\d \utv_i \over \d x_j}  {\d \utv_j \over \d x_i}$,
and we can eliminate it by setting
$$ p = \sum\limits_{i,j} {\d \utv_i \over \d x_j}  {\d \utv_j \over \d x_i} .
\eqno{(2.9)}$$
Now the first two equations in (2.8) reduce to a single equation
$${\d \tv \over \d t} =  (\B_0 \cdot \nabla) \tv . \eqno{(2.10)} $$
Finally, by taking an arbitrary divergence zero vector field $\w (x)$,
we can construct a solution of (2.8):
$$\tv(x,t) = \w (x + \B_0 t) .$$
This traveling wave solution is called the \Alfven wave. The only difference
with the classical MHD system (2.1) is the change of the pressure term (2.9).

\

\

{\bf 3. Generalized Euler equation.}

\

 In this section we are going to review the geodesic equation approach to hydrodynamics
developed by Arnold. In the key paper [2], Arnold gave an interpretation of the
Euler equation for an incompressible ideal fluid  
$$\left\{
\matrix{
{\d \bv \over \d t} = - (\bv \cdot \nabla) \bv  - \nabla p \hfill\cr
\div (\bv) = 0, \hfill\cr
}
\right.
\eqno{(3.1)}
$$
from the perspective of infinite-dimensional Lie groups. He showed that the Euler
equation may be interpreted as the geodesic equation on the group of diffeomorphisms,
where the Riemannian structure on the group is given by the energy functional.

 We will describe this approach here following the book [3].

 Let $G$ be a Lie group (possibly infinite-dimensional), and let $\g$ be its Lie algebra.
Consider a map from $\g$ to its dual
$$ A: \, \g \rightarrow \g^* 
\hbox{\rm \ \ \ (inertia operator),} $$
such that it defines a positive-definite symmetric bilinear form on $\g$:
$$ <X|Y> = {1\over 2} A(X) Y + {1\over 2} A(Y)X , \; \; \; X,Y \in \g . $$
The corresponding quadratic form $<X|X> = A(X) X$ is called the energy functional.

 The Lie algebra $\g$ acts on its dual space via the coadjoint action:
$$ \left( \ad^*(X) u \right) (Y) = - u \left( [X,Y] \right)
\; \; \; {\rm for} \;  X,Y \in \g, \, u \in \g^* .$$
We assume that the space $A(\g)$ is invariant under the coadjoint action, and we make 
a convention that in what follows $\g^*$ stands for $A(\g)$ (this is a slight abuse
of notations since in the infinite-dimensional case $A(\g)$ is typically smaller
than the formal dual of $\g$). Since the kernel of $A$ is trivial (otherwise,
the quadratic form $<X|X> = A(X) X$ is not positive-definite), then with the above
convention the operator $A$ is invertible.

 The bilinear form $<\cdot|\cdot>$ may be left-translated from $\g = T_e (G)$ to
the tangent spaces at all points of $G$. This gives a Riemannian structure
on $G$, and allows us to consider the geodesics on this group.

 Next we are going to write the equation for the geodesics on $G$, which describes
the evolution of the tangent vector $X \in \g$ to the geodesic curve. It turns out
however, that it is easier to write the evolution equation for the covector
$u = A(X) \in \g^*$ rather than for $X$ itself. The generalized Euler equation is
the evolution equation for $u = A(X)$ which is written using the coadjoint action
(see (6.4) in [3]):
$$ u_t = - \ad^* \left(A^{-1} u \right) u . \eqno{(3.2)} $$

When $G = SO(3)$ this equation turns into the equations of motion of a rigid body
with a fixed point.

Let us discuss equation (3.2) in the context of fluid dynamics. Evolution of an
incompressible fluid in domain $D\subset \R^n$ from time $0$ to time $t$ is
given by a volume-preserving diffeomorphism of $D$. Thus the group of the volume preserving 
diffeomorphisms $G = \SD(D)$ is the configuration space for this motion. The Lie algebra
of the group $\SDD$ is the Lie algebra of the divergence zero vector fields $\SVD$.
As the energy functional we take the kinetic energy:
$$\left< \sum_i v_i (x) {\d \over \d x_i} \big| \sum_i v_i (x) {\d \over \d x_i} \right>
= \int_D \sum_i v_i (x)^2 dV .
\eqno{(3.3)}
$$  

 The dual space for the divergence zero vector fields is the factor space of
differential 1-forms on $D$ by exact 1-forms:
$$\SVD^* = \KD .$$
The pairing between $\SVD$ and $\KD$ is given by the integral:
$$\sum\limits_j w_j (x) dx_j \left( \sum\limits_i v_i (x) {\d \over \d x_i} \right)
= \int_D \sum\limits_j w_j(x) v_j(x) dV. \eqno{(3.4)}$$
It is easy to check that exact 1-forms vanish on the divergence zero vector 
fields, 
and so the value of the integral on the right is independent of the choice of a
representative in a class of 1-forms modulo $d \Omega^0 (D)$.

The inertia operator
$$ A: \; \SVD \rightarrow \KD $$
that corresponds to the energy functional (3.3) is written as follows:
$$A \left( \sum\limits_i v_i (x) {\d \over \d x_i} \right) =
\sum\limits_i v_i (x) d x_i .$$
It is possible to check that in these settings, equation (3.2) turns into the Euler     
equations of motion of an ideal fluid.

The equations of the magnetic hydrodynamics (2.1) can also be obtained as 
a special case of the generalized Euler equation (3.2). To construct the 
corresponding Lie algebra we take the Lie algebra $\SVD$ together with 
its dual space:
$$\g = \SVD \oplus \KD .$$
The Lie bracket of two 1-forms is set to be zero. The Lie bracket of 
two vector fields is defined in the usual way:
$$\left[ \sum\limits_i v_i(x) {\d \over \d x_i} , 
\sum\limits_j w_j(x) {\d \over \d x_j} \right]
= \sum\limits_{i,j} v_i {\d w_j \over \d x_i} {\d \over \d x_j}
- w_j {\d v_i \over \d x_j} {\d \over \d x_i} , \eqno{(3.5)}
$$
and the Lie bracket of a vector field with a 1-form is given by the
Lie derivative:
$$\left[ \sum\limits_i v_i(x) {\d \over \d x_i} , 
\sum\limits_j w_j(x) d x_j \right]
= \sum\limits_{i,j} v_i {\d w_j \over \d x_i} d x_j
+ \sum\limits_j w_j d(v_j)  . \eqno{(3.6)}
$$
It is easy to see that the space $d\Omega^0 (D)$ is invariant under
the Lie derivative action, so the above formula may be taken modulo
$d\Omega^0 (D)$.

The Lie algebra $\g$ is a semidirect product of $\SVD$ with its module
$\KD$. The space $\KD$ forms an abelian ideal in $\g$.

A really important feature of the Lie algebra $\g$ is the existence of
an invariant symmetric non-degenerate bilinear form. In contrast,
the Lie algebra $\SV$ does not possess such a form. The invariant form on
$\g$ is defined as follows (cf. (3.4)):
$$\matrix{
\left( \sum\limits_i v_i (x) {\d \over \d x_i} \bigg|
\sum\limits_j w_j (x) dx_j  \right)
= \int_D \sum\limits_i v_i(x) w_i(x)  dV, \cr
\left( \sum\limits_i v_i (x) {\d \over \d x_i} \bigg|
\sum\limits_j w_j (x) {\d \over \d x_j}  \right) = 0,
\;
\left( \sum\limits_i v_i (x) dx_i \bigg|
\sum\limits_j w_j (x) dx_j  \right) = 0. \cr
}
\eqno{(3.7)}$$
One can verify that the bilinear form (3.7) satisfies the invariance 
property:
$$ \left( \left[ X, Y \right] \big | Z \right)
= \left( X  \big | \left[Y , Z \right] \right)
\; \; \hbox{\rm for all \ } X, Y, Z \in \g. $$

 We can use this form to identify each element $X\in \g$ with a linear
functional $(X | \cdot )$ in $\g^*$. It is well-known that when the bilinear
form that is used to identify $\g^*$ with $\g$, is invariant and 
non-degenerate, the coadjoint action becomes isomorphic to the
adjoint action. In this case the generalized Euler equation takes
form:
$$ X_t = - \ad (A^{-1} X) X, \eqno{(3.8)}$$
where now $X\in \g$ and the inertia operator $A$ now maps $\g$ to $\g$.

 The generalized Euler equation (3.8) yields the equations of magnetic
hydrodynamics (2.1) if we choose $A$ to be the following involution
on $\g$:
$$\matrix{
A \left( \sum\limits_i v_i (x) {\d \over \d x_i} \right) =
\sum\limits_i v_i (x) d x_i , \cr
A \left( \sum\limits_i w_i (x) d x_i \right) =
\sum\limits_i w_i (x) {\d \over \d x_i} . \cr
}
\eqno{(3.9)}
$$
Note that in the second  equality we choose a (unique) representative
in a class modulo $d \Omega^0$ satisfying 
$\sum\limits_i {\d w_i  \over \d x_i} = 0$,
so that the right hand side is a divergence zero vector field.

 We see that the inertia operator (3.9) satisfies $A^{-1} = A$,
and the energy functional $(AX | X)$ for
$X = \sum\limits_i v_i (x) {\d \over \d x_i}
+ \sum\limits_i w_i (x) d x_i \in \g$ 
is given by the integral
$$ \int_D \sum\limits_i v_i (x)^2 + \sum\limits_i w_i (x)^2 dV .
\eqno{(3.10)}
$$

The Lie algebra 
$$ \g = \SV \oplus \K$$
appears in the study of toroidal Lie algebras ([4], [5], [6], [9], [11]). 
The representations of $\g$ are an essential ingredient for 
constructing the representation theory of toroidal Lie algebras.
It was discovered however that $\g$ admits a non-trivial deformation with a
$\K$-valued 2-cocycle on $\SV$, and one gets a better representation
theory for the deformed algebra than for $\g$ itself.

In the next Section we will describe this deformation of $\g$ and study
the associated generalized Euler equation.

\

\

{\bf 4. Abelian extensions of the Lie algebra of vector fields.}

\

In the previous Section we have constructed a semidirect product $\g$
of the Lie algebra of divergence zero vector fields with $\K$.
It turns out that on the same vector space
$$ \g = \SV \oplus \K$$
we may deform the Lie bracket in a non-trivial way. When we define the
Lie bracket of two vector fields, we are going to add to the right hand 
side of (3.5) a correction term which has value in $\KD$:
$$\left[ \sum\limits_i v_i(x) {\d \over \d x_i} , 
\sum\limits_j w_j(x) {\d \over \d x_j} \right]
$$
$$
= \sum\limits_{i,j} \left( v_i {\d w_j \over \d x_i} {\d \over \d x_j}
- w_j {\d v_i \over \d x_j} {\d \over \d x_i} \right) 
+ \tau \left( \sum\limits_i v_i(x) {\d \over \d x_i} , 
\sum\limits_j w_j(x) {\d \over \d x_j} \right). 
\eqno{(4.1)}
$$
In order to get a Lie bracket, $\tau$ has to be a 2-cocycle on $\SVD$
with values in  \break
$\KD$.
% $\tau\in H^2 \left(\SVD,\KD \right)$.

The following cocycle plays an important role in the representation
theory:
$$ \tau \left( \sum\limits_i v_i(x) {\d \over \d x_i} , 
\sum\limits_j w_j(x) {\d \over \d x_j} \right) 
= \sum\limits_{i,j} {\d v_i \over \d x_j} d \left( {\d w_j \over \d x_i} \right).
\eqno{(4.2)}$$

This cocycle may be viewed as a higher-dimensional generalization of the Virasoro 
cocycle. Just as the Virasoro cocycle, it has a triple derivative in $x$, and in fact 
(4.2) reduces to the Virasoro cocycle for the Lie algebra of vector fields on a circle.

We will denote the Lie algebra with the Lie bracket deformed by the cocycle $\tau$
by $\g(\tau)$. Note that
$$\g(\tau) = \SVD \oplus \KD$$
is no longer a semidirect product, but the subspace $\KD$ still forms an abelian ideal.
As before, the action of $\SVD$ on $\KD$ is given by the Lie derivative formula (3.6). 

\

{\bf Proposition 4.} The bilinear form on $\g(\tau)$ given by (3.7) is invariant.

\noindent
{\sl Proof.} We need to establish the invariance property:
$$ \left( \left[ X, Y \right] \big | Z \right)
= \left( X  \big | \left[Y , Z \right] \right) .$$
There are three non-trivial cases to be considered:

\noindent
(i) $X, Y \in\SVD, Z \in \KD$,

\noindent
(ii) $X, Z \in\SVD, Y \in \KD$,

\noindent
(iii) $X, Y, Z \in\SVD$.

We will verify the invariance only for the last case, since only this case will involve the cocycle $\tau$, and leave the first two cases as an 
exercise to the reader.

Suppose $X = \mathop\sum\limits_i u_i {\d\over \d x_i}$,
$Y = \mathop\sum\limits_j v_j {\d\over \d x_j}$,
$Z = \mathop\sum\limits_k w_k {\d\over \d x_k}$.
Since $\left( \SV \big| \SV \right) = 0$, then we get that
$ \left( \left[ X, Y \right] \big | Z \right)
=  \left( \tau \left( X, Y \right) \big | Z \right)$
and 
$\left( X  \big | \left[Y , Z \right] \right) =
\left( X  \big | \tau \left(Y , Z \right) \right)$.

We have
$$ \left( \left[ X, Y \right] \big | Z \right)
= \left(
\sum\limits_{i,j,s} {\d u_i \over \d x_j} {\d^2 v_j \over \d x_i \d x_s} 
d x_s 
\big| \sum\limits_k w_k {\d\over \d x_k} \right) $$
$$ = \int_D \sum\limits_{i,j,k} w_k {\d u_i \over \d x_j} 
{\d^2 v_j \over \d x_i \d x_k}  dV .$$
Integrating by parts and using the fact that 
$\mathop\sum\limits_i {\d u_i \over \d x_i} = 0$, we get
$$\left( \left[ X, Y \right] \big | Z \right)
= - \int_D \sum\limits_{i,j,k} 
{\d w_k \over \d x_i} {\d u_i \over \d x_j} 
{\d v_j \over \d x_k}  dV .$$

On the other hand,
$$\left( X  \big | \tau \left(Y , Z \right) \right)
= \left( \sum\limits_i u_i {\d\over \d x_i} \big|
\sum\limits_{j,k,s} {\d v_j \over \d x_k} {\d^2 w_k \over \d x_j d x_s} 
d x_s \right) $$
$$ = \int_D \sum\limits_{i,j,k} u_i 
{\d v_j \over \d x_k} {\d^2 w_k \over \d x_j d x_i} dV
= - \int_D \sum\limits_{i,j,k} 
 {\d u_i \over \d x_j} {\d v_j \over \d x_k} {\d w_k \over \d x_i} dV .$$
This proves the invariance property 
$ \left( \left[ X, Y \right] \big | Z \right)
= \left( X  \big | \left[Y , Z \right] \right)$ in case (iii).

\

Now we are going to prove the following

{\bf Theorem 5.} The generalized Euler equation 
$$ X_t = - \left[ AX, X \right] \eqno{(4.3)}$$ 
for the Lie algebra $\g(\tau)$ with the inertia operator $A$
given by (3.9) yields the equations of magnetic hydrodynamics with
asymmetric stress tensor (2.2).
 
\noindent
{\sl Proof.} We write $X = \mathop\sum\limits_i B_i {\d \over \d x_i}
+  \mathop\sum\limits_j v_j d x_j $. We will fix representatives of classes
of 1-forms modulo $d \Omega^0 (D)$ by imposing a condition
$\mathop\sum\limits_j {\d v_j \over \d x_j} = 0$. Then we have
$$ AX = \sum\limits_j v_j {\d \over \d x_j} + \sum\limits_i B_i d x_i $$
and
$$ \left[ X, AX \right] = 
 \sum\limits_{i,j} B_i {\d v_j \over \d x_i} {\d \over \d x_j}
- \sum\limits_{i,j} v_j {\d B_i \over \d x_j} {\d \over \d x_i}
+ \sum\limits_{i,j,k} 
{\d B_i \over \d x_j} {\d^2 v_j \over \d x_i \d x_k} d x_k
$$
$$
+ \sum\limits_{i,j} B_i {\d B_j \over \d x_i} d x_j
+ \sum\limits_{i} B_i d (B_i) 
- \sum\limits_{i,j} v_i {\d v_j \over \d x_i} d x_j
- \sum\limits_{i} v_i d (v_i). 
$$
Note that the terms 
$B_i d(B_i) = {1\over 2} d(B_i^2)$ 
and
$v_i d(v_i) = {1\over 2} d(v_i^2)$ 
are full differentials and thus may be
dropped.

 Substituting the obtained expression into the generalized Euler equation (4.3)
and collecting terms at $d x_j, {\d \over \d x_j}$, 
and taking into account that equality of 1-forms is taken modulo $d \Omega^0 
(D)$,
we get the following
system of PDEs:
$${\d v_j \over \d t} = - \sum\limits_i v_i {\d v_j \over \d x_i}
+ \sum\limits_i B_i {\d B_j \over \d x_i} 
+ \sum\limits_{i,k} 
{\d B_i \over \d x_k} {\d^2 v_k \over \d x_i \d x_j}
- {\d p \over \d x_j} ,$$
$${\d B_j \over \d t} = - \sum\limits_i \left( v_i {\d B_j \over \d x_i}
- B_i {\d v_j \over \d x_i} \right) , $$
$$ \sum\limits_j {\d v_j \over \d x_j} = 0, \; 
\sum\limits_j {\d B_j \over \d x_j} = 0. $$
Rewriting this system in a vector form with the vector fields
% $\bv = (v_1(x), \ldots, v_n(x)), \B = (B_1(x), \ldots, B_n (x))$, 
we get (2.2).
% $$\left\{
% \matrix{
% {\d \bv \over \d t} = - (\bv \cdot \nabla) \bv + (\B \cdot \nabla) \B 
% - \sum\limits_{i,j} {\d B_i \over \d x_j} \nabla {\d v_j \over \d x_i} 
% - \nabla p \hfill\cr
% {\d\B \over \d t} = - \left\{ \bv, \B \right\} \hfill\cr
% \div (\bv) = 0, \, \, \div (\B) = 0. \hfill\cr
% }
% \right.
% $$

Finally let us prove Theorem 3 and establish the energy 
and the cross-helicity conservation laws
for MHD equations with the stress tensor (2.2).
We will in fact obtain Theorem 3 as a corollary of the following general

{\bf Theorem 6.} Let $\g$ be a Lie algebra with a non-degenerate symmetric
invariant bilinear form $(\cdot | \cdot)$. Let $A$ be an involution of $\g$
preserving the invariant form,
$$A: \; \g \rightarrow \g, \; \; A^2 = {\rm Id}, \; \; 
\left( AX | AY \right) = \left( X | Y \right) \; \hbox{\rm for all \ }
X,Y\in \g .$$
Then the generalized Euler equation
$ X_t = - \left[ AX, X \right] $
has the following two first integrals:
$$ \left( AX | X \right) \equiv Const  \eqno{(4.4)}$$
and 
$$ \left( X | X \right) \equiv Const.  \eqno{(4.5)}$$

\noindent
{\sl Proof.}
Let us evaluate ${\d \over \d t} \left( AX | X \right)$:
$${\d \over \d t} \left( AX | X \right) =
\left( AX_t | X \right) + \left( AX | X_t \right) .$$
Taking into account that $(X|Y) = (AX | AY)$ and $A^2 = \Id$, we get that
$( AX_t | X ) = (A^2 X_t | AX ) = (X_t | AX)$. Thus
$${\d \over \d t} \left( AX | X \right) = 2 \left( AX | X_t \right) .$$
Substituting the right hand side of the generalized Euler equation for $X_t$
we obtain
$${\d \over \d t} \left( AX | X \right) = 
- 2 \left( AX | \left[AX , X \right] \right) .$$
By invariance of the form we get
$$ \left( AX | \left[AX , X \right] \right) = 
\left( \left[AX , AX \right]  |  X \right) = 0 .$$
Thus  ${\d \over \d t} \left( AX | X \right) = 0 $ and (4.4) is established.

The second conservation law (4.5) is obtained in a similar way:
$${\d \over \d t} \left( X | X \right) = 
2 \left( X_t | X \right) = - 2 \left(  \left[AX , X \right] | X \right)
= - 2 \left(  AX  | \left[ X , X \right] \right) = 0 .$$
This completes the proof of Theorem 6. 

Note that for (4.5) we may drop the requirements that $A$ preserves
the invariant form and is an involution. 

We obtain Theorem 3 as an immediate corollary to the previous Theorem, noting that
the bilinear form (3.7) on $\g(\tau)$ is invariant by Proposition 4 and
the inertia operator (3.9) is an involution and preserves this form.

We can see that for $X = \mathop\sum\limits_i B_i {\d \over \d x_i}
+  \mathop\sum\limits_j v_j d x_j $, the first integral (4.4) becomes the
energy conservation law:
$$\left( AX | X \right) = 
\int_D \sum_i v_i(x)^2 + \sum_i B_i(x)^2 dV \equiv Const, $$
and (4.5) becomes the cross-helicity conservation:
$$\left( X | X \right) = 
2 \int_D \sum_i v_i(x) B_i(x) dV \equiv Const .$$

\

\

{\bf Acknowledgment.} I thank Boris Khesin for getting me interested in this problem,
and David Amundsen for helpful discussions. 
I have benefited from the hospitality of l'Institut des Math\'ematiques
de Jussieu (Paris), where part of this work has been done. This research is supported
by the Natural Sciences and Engineering Research Council of Canada.

\

\

{\bf References:}

\noindent
[1] Adams, R.A., Sobolev spaces, Pure and applied mathematics, v.65,
Academic Press, N.Y., 1975.

\noindent
[2] Arnold, V., ``Sur la g\'eom\'etrie diff\'erentielle des groupes de Lie de 
dimension 
infinie et ses applications \`a l'hydrodynamique des fluides parfaits'', 
Ann. Inst. Fourier (Grenoble), {\bf 16}, 319-361, 1966.

\noindent
[3] Arnold, V.I., Khesin, B.A., Topological methods in hydrodynamics,
Applied Mathematical Sciences, v.125, Springer-Verlag, N.Y., 1998.

\noindent
[4] {Berman, S. and Billig, Y.},
``Irreducible representations for toroidal Lie algebras'',
{ J. Algebra}, {\bf 221}, 188--231, 1999.

\noindent
[5] {Billig, Y.},
``Principal vertex operator representations for toroidal Lie 
algebras'',
{J. Math. Phys.}, {\bf 39}, 3844--3864, 1998.

\noindent
[6] {Billig, Y.},
{``Energy-momentum tensor for the toroidal Lie algebras''},
preprint, 
\hfill\break
math.RT/0201313.

\noindent
[7] {Billig, Y.},
{``Abelian extensions of the group of diffeomorphisms of a torus''},
{Lett. Math. Phys.}, {\bf 64}, 155--169, 2003.

\noindent
[8] Biskamp, D., Nonlinear magnetohydrodynamics, Cambridge University Press, 
Cambridge, 1993.

\noindent
[9] {Eswara Rao, S. and Moody, R.V.}:
{``Vertex representations for $N$-toroidal Lie algebras and a 
generalization of the Virasoro algebra'',}
{ Comm. Math. Phys.}, {\bf 159}, 239--264, 1994.

\noindent
[10] Kadomtsev, B.B., Tokamak plasma: a complex physical system, Institute of Physics
Publishing, Bristol, 1992.

\noindent
[11] {Larsson, T.A.}:
{``Lowest-energy representations of non-centrally extended diffeomorphism algebras'',}
{ Comm. Math. Phys.}, {\bf 201}, 461--470, 1999.

\noindent
[12] Marsden, J., Ratiu, T., Weinstein, A., 
``Semidirect products and reduction in mechanics'',
Trans. Amer. Math. Soc., {\bf 281}, 147--177, 1984.

\noindent
[13] Moffatt, H.K., ``Vortex- and magneto-dynamics -- a topological perspective'',
Mathematical Physics 2000, 170-182, Imp. Coll. Press, London, 2000.

\noindent
[14] Moffatt, H.K., ``Some developments in the theory of turbulence'',
J. Fluid Mech., {\bf 106}, 27-47, 1981.

\noindent
[15] Moreau, R., Magnetohydrodynamics, Fluid mechanics and its applications, v. 3,
Kluwer Academic Publishers, Dodrecht, 1990.

\noindent
[16] Ovsienko, V.Yu., Khesin, B.A., 
``The super Korteweg-de Vries equation as an Euler equation'',
Funktsional. Anal. i Prilozhen., {\bf 21}, 81-82, 1987.

\noindent
[17] Priest, E.R., Solar magnetohydrodynamics, Geophysics and astrophysics 
monographs, v. 21, D. Reidel Publishing, Dodrecht, 1982.

\noindent
[18] Rosensweig, R.E., Ferrohydrodynamics, Cambridge University Press, Cambridge, 
\break
1985.

\noindent
[19] Vi{\v{s}}ik, S.M., Dol{\v z}anski{\u{i}}, F.V., ``Analogs of the 
Euler-Lagrange
equations and magnetohydrodynamics connected with Lie groups'',
Dokl. Akad. Nauk SSSR, {\bf 19}, 149-153, 1978.

\end